\documentclass[prd,aps,eqsecnum,psfig,showpacs]{revtex4}
\begin{document}
\title{Braneworld gravitational collapse from a radiative bulk}
\author{Supratik Pal \footnote{Electronic address: {\em supratik@cts.iitkgp.ernet.in}}}
\affiliation{Centre for Theoretical Studies \\
Indian Institute of Technology \\ Kharagpur 721 302, India \\
and \\
Relativity and Cosmology Research Centre \\ Department of Physics \\
Jadavpur University \\  Kolkata 700 032, India}
\vspace{.5in}

\begin{abstract}

We study the fate of a collapsing star on the brane in a generalized
braneworld gravity with bulk matter. Specifically,
we investigate for the possibility of having a static exterior for a collapsing
star in the radiative bulk scenario. Here, 
the nonlocal correction due to bulk matter  is manifest in an induced mass 
that adds up to the physical mass of the star resulting in an effective mass.
 A Schwarzschild solution for the 
exterior in terms of this effective mass is obtained, which reveals that
even if the star exchanges energy with the bulk, the
exterior may appear to be static to a braneworld observer located outside the
collapsing region. The possible explanation of the situation 
from the discussion on the role of bulk matter is provided.
The nature of bulk matter and the corresponding bulk geometry have also 
been obtained and analyzed, which gives a complete picture of both brane and bulk
viewpoints.

\end{abstract}


\maketitle


\section{Introduction}

The search for a consistent description of black hole physics and gravitational collapse
 in braneworlds \cite{rs2} has been a challenge to theoretical physics.
As a matter of fact, black holes and gravitational collapse are not yet
 well-understood in the braneworld scenario. The idea of an extended
singularity in higher dimensions leads to a black string solution \cite{blstring}
which is neither stable nor localized on the brane. The first solution for
localized black holes on the brane came out to be
Reissner-N\"{o}rdstrom type with a `tidal charge' contribution arising from the bulk Weyl
tensor \cite{tidal}.  Later on, attempts were made to include a Schwarzschild metric 
with non-vacuum brane and the black hole intersecting the bulk 
\cite{seahra, adsbh}. Solutions for the charged rotating black holes on the brane was 
also obtained \cite{chargerot}. (see also \cite{revbh} for a review and \cite{bhrecent} for 
some very recent results). 
Based on this tidal charge scenario, Oppenheimer-Snyder type \cite{os}
gravitational collapse of spherically symmetric objects 
was studied in \cite{coll1} that
led to an interesting conclusion. This was formulated by a no-go theorem
that indicates  a non-static exterior for
the collapsing sphere on the brane. Subsequently, it was shown  
that the exterior for this radiative sphere can be described by a
Vaidya  metric that envelops the  collapsing region \cite{coll2}. 
Possible  generalizations of the non-static nature for induced gravity with or
without the Gauss-Bonnet term are also around \cite{collgen}.
However, it was demonstrated in \cite{coll3} that  a static exterior can be obtained 
 by relaxing the idea of dust inside the star, thereby introducing a non-vanishing surface pressure, 
and by ignoring the tidal effect.

Out of simplicity, these descriptions are based on the
assumption that the bulk is empty,  comprising of a negative
cosmological constant. No doubt, in the simplest case, they
provide important insight to the bulk-brane interplay.
But a careful look at the scenario reveals that the source of the above conclusions 
may, in fact, lie in this very assumption. Nevertheless, an empty bulk might be only a
 special situation to deal with. 
For a more realistic description of the physics on the brane, it is  instructive 
 to take into account the effects of bulk fields as well.
There is many a reason  why one should consider bulk fields.
The  braneworld models are motivated by the $p$-brane solutions of 
M theory \cite{bulkst}, which are obtained by considering 
scalars (e.g., dilaton), 3-form fields (e.g., Kalb-Ramond field) etc. in the bulk.
It is  expected that these properties of  $p$-branes will be reflected
in  braneworld physics too. 
Beside  phenomenological motivations \cite{bulkph},
the urge of considering bulk fields comes from the so-called radion stabilization problem,
which was solved by the Goldberger-Wise mechanism \cite{goldwise}.
From the geometric point of view, a bulk field can also be thought of as an 
outcome of global topological defects which supply a non-trivial stress-energy
tensor outside the brane \cite{topo}.
The motivation from the gravitational sector include addressing the age-old
problems like the cosmological constant problem \cite{coscon}. 
Also, a Schwarzschild black hole
on the brane will have a regular AdS horizon 
only if the bulk contains exotic matter \cite{adshor}.
In fact, in a realistic brane cosmological scenario, 
the bulk gravitons produced by the fluctuations on brane matter  act
like an effective field residing in the bulk \cite{langrev}. 
Thus the need for considering matter fields in the bulk is apparently unavoidable.
With this motivation, we intend to
analyze gravitational collapse for a  non-empty bulk.

It has been shown extensively in numerous papers \cite{maartbulk, lang1, langrev, chamb}
that  in presence of bulk matter, the bulk metric
for which an FRW geometry on the brane is recovered, is given by a 
higher dimensional  generalization  of the  Vaidya AdS (VAdS$_{5}$)  spacetime
 \cite{vaid}, with the radiation flowing from the brane to the bulk (incoming radiation). 
This black hole, being radiative, exchanges energy with the brane,
which is manifest through a \textit{brane
matter non-conservation equation}, so that the matter on the brane is no longer
strictly conserved \cite{maartrev}. However, the total matter-energy of the bulk-brane system 
still remains conserved, confirming no global violation of matter conservation. 
Of late, the idea of the VAdS$_{5}$  radiative bulk scenario has been
generalized for both incoming and outgoing radiation \cite{vads5gen}.
This generalized idea has opened up new avenues of visualizing the brane phenomena 
from the point of view of bulk-brane energy exchange. 
Subsequently, different possibilities have been investigated.
 For example, \cite{asymgen} discusses the scenario for asymmetric embedding and
\cite{nonradial} accounts for non-radial emission of bulk gravitons. Several other models
with radiating black holes in the bulk have also been brought forth \cite{radbh5}.

Our aim in this article is to utilize the generic feature of the VAdS$_{5}$ bulk discussed in \cite{vads5gen}
in order  to study gravitational collapse of spherically symmetric objects. 
In our model, the bulk energy-momentum tensor is given by a  phantom (ghost)
radiation field having negative energy density. The phantom field plays an important role in 
gravitational physics. In four dimensions,
 it raises a fair possibility of explaining observed accelerated expansion of the universe
\cite{phant1} and  unifying dark mater and dark energy \cite{phant2}. 
Additionally, the radiative behavior of a phantom null dust have been employed to obtain
wormhole  solutions  \cite{phantworm}. 
In the braneworld context too, a phantom field in the bulk has shown much promise.
It has been shown in \cite{rkskphant} that a bulk phantom field 
makes it possible to localize both massless and massive fermions on the brane
and in six dimensional models, it results in the localization of massless modes of all the standard model 
fields as well as gravity on the same brane \cite{rksk6d}. Nevertheless, since such exotic matter fields
can indeed be present in the bulk, it is interesting to investigate for the possible consequences 
of these fields on the physics 
of the four dimensional world. This serves as the basic motivation for considering phantom
radiation in the bulk and study gravitational collapse on the brane.
We show, with the help of the modified Einstein equation and the  non-conservation equation, that the 
collapsing star on the brane can exchange energy with the bulk even if
the exterior may appear to be static. 
The unique feature of this radiative bulk scenario  is that a braneworld (local)
observer located near the surface of the star cannot {\em feel} this effect
so that the star would appear to be  static even though there is energy-flow between the brane and the bulk.
In this way, we generalize the 
analysis of gravitational collapse of spherical objects on the brane, and at the same
time, re-establish the possibility of having a static exterior for a collapsing sphere.

Throughout the article, we use $\mu,\nu,...$ for the brane indices and $M,N,...$
for those in the bulk. Specifically, 
we choose the following notations for the coordinates : $(\tau, ~r, ~\theta, ~\phi) \equiv
$ brane coordinates for the interior of the collapsing sphere; $(T, ~R, ~\Theta, ~\Phi) \equiv$ brane
coordinates for the exterior; and $(t, ~{\cal R}, ~x, ~y, ~z)\equiv$ bulk coordinates 
$\equiv (v, ~{\cal R}, ~x, ~y, ~z)$ in terms of the null coordinates.


\section{Brane viewpoint with a radiative bulk}

In presence of a bulk field exchanging energy with the brane, the  effective Einstein equation
on the brane \cite{eee} is  generalized to \cite{maartbulk} 
\begin{equation}
G_{\mu\nu} = - \Lambda g_{\mu\nu} + \kappa^2 T_{\mu\nu} +
\kappa_5^4 {\cal S}_{\mu\nu} - {\cal E}_{\mu\nu} + {\cal F}_{\mu\nu}
\label{eee}
\end{equation}
where $\mu,\nu,...$ are the brane (4D) indices; ${\cal S}_{\mu\nu}$, ${\cal E}_{\mu\nu}$ and ${\cal F}_{\mu\nu}$
are respectively the quadratic  brane energy-momentum tensor,
the projected bulk Weyl tensor and the  bulk energy-momentum
tensor projected on the brane and $\kappa_5^4 = 6 \kappa^2/\lambda$
the 5D coupling constant. For a general VAdS$_{5}$ bulk with both the possibilities of
incoming and outgoing radiation, 
the matter conservation equation on the brane is  modified to \cite{vads5gen}
\begin{equation}
\dot\rho + 3 \frac{\dot a}{a} (\rho + p) = - 2 \epsilon \psi
\label{rho} 
\end{equation} 
where $\epsilon \psi \propto d {\cal M} /d v$, with ${\cal M}(v)$ the sumtotal of the masses of
the bulk black hole and the radiation field. A crucial term in the above expression is
 $\epsilon$ which can take the values $\epsilon = \pm 1$ and
 results in the bulk-brane energy-exchange. Since for a phantom radiative field, $\psi <0$, 
it appears that $\epsilon < 0 \Rightarrow d {\cal M} /d v > 0$ so that the bulk gains energy
whereas $\epsilon > 0 \Rightarrow d {\cal M} /d v < 0$ implying that the bulk loses energy.
A negative signature for $\epsilon$ will thus indicate that an object on the brane releases phantom radiation
 to the bulk whereas a positive $\epsilon$ will mean that it absorbs phantom radiation 
from the bulk. To a braneworld observer,  $\psi$ is the quantitative
estimate of the brane-projection of the bulk energy density, reflected
by the equation 
\begin{equation}
{\cal F}_{\mu\nu} = \frac{2}{3}\kappa_5^2 ~ \psi h_{\mu\nu}
\label{fmunu} 
\end{equation}
where $h_{\mu\nu}$ is the induced metric on the brane.  
 Consequently, the Bianchi identity on the brane  leads to the equation
governing the evolution of the so-called Weyl term
\begin{equation}
\dot\rho^* + 4 \frac{\dot a}{a} \rho^* = 2 \epsilon \psi - \frac{2 \kappa_5^2}
{3\kappa^2} \left[\dot \psi + 3 \frac{\dot a}{a} \psi \right] 
\label{eqrho*} 
\end{equation}
Equations (\ref{rho}) and (\ref{eqrho*}) show that in general there is 
a coupling  \cite{struc} between the bulk
energy-momentum tensor and its brane counterpart, that is responsible for the
bulk-brane energy-exchange. However, as already pointed out in \cite{struc},
since the bulk matter is not determined 
{\em a priori}, the ansatz for the coupling term ${\cal Q}$ (which is precisely
the right hand side of the above equation) can be arbitrarily chosen,
so far as it is physically acceptable. 
Several possible ansatz have been considered in \cite{maeda, excos1, excos2, excos3, struc}
and the  consequences have been investigated. Brane cosmological dynamics of
bulk scalar fields have been studied in detail in \cite{dynscal, langrev, effmass}. 
What follows is that we can  take an ansatz for the coupling term of the form
${\cal Q} = H \rho^*$.
With this  ansatz, Eq (\ref{eqrho*}) reveals that the Weyl term behaves as 
\begin{equation}
\rho^* = \frac{C(\tau)}{a^4}, ~~~C(\tau) = C^* a(\tau)
\label{rho*}
\end{equation}
where $C(\tau)$ is the scaled on-brane mass function  ${\cal M}(\tau)$,
 $\tau$ being the proper time on the brane.
Hence, for this type of ansatz, the Weyl term supplies an additional matter-like effect to
the brane. We shall show \textit{a posteriori} what type of bulk matter can
give rise to  this matter-like nature of the Weyl term.


\section{Gravitational collapse on the brane}

Let us now analyze gravitational collapse for spherically symmetric objects
using the effective Einstein equation with bulk fields discussed above.
For a sphere undergoing Oppenheimer-Snyder collapse, 
the collapsing region can be conveniently expressed  by a Robertson-Walker metric
\begin{equation}
ds^2_{\text{int}} =-d\tau^2 + a^2(\tau) (1+ {\textstyle {1\over4}}
kr^2)^{-2}\left[dr^2+r^2d\Omega^2\right]
\label{int}
\end{equation}
with $(\tau, ~r, ~\theta, ~\phi)$ as the brane coordinates for the interior.
This metric  has to be a solution of the generalized
brane Friedmann equation, which, 
with the the help of Eq (\ref{rho*}) and the RS fine-tuning
($\Lambda = 0$), turns out to be
\begin{equation} 
\left( \frac{\dot a}{a} \right)^2 =  \frac{8 \pi G}{3}   \rho \left(1 +
\frac{\rho}{2\lambda} \right) + \frac{C^*}{ a^3} - \frac{k}{a^2}
\label{fridint}
\end{equation}
Also, the Raychaudhuri equation for geodesic focusing can  be written as
\begin{equation}
\frac{\ddot a}{a} = -  \frac{4 \pi G}{3} \left[\rho \left(1 +\frac{\rho}{2\lambda} \right)
+ 3 p \left(1 +\frac{\rho}{\lambda} \right) \right] - \frac{C^*}{ a^3}
-\frac{\kappa_5^2}{3} \psi
\label{rc}
\end{equation}
Now, expressing in terms of the proper radius 
 $R(\tau)=r a(\tau)/(1+ {\textstyle{1\over4}} kr^2)$, 
the generalized Friedmann equation (\ref{fridint}) for the collapsing region reads
\begin{equation}
\dot{R}^2= \frac{2GM}{R}+ \frac{3 G M^2}{4\pi\lambda R^4}+ \frac{2 G M^*} {R} + E
\label{frid}
\end{equation}
with
\begin{equation}
M = \frac{4}{3}\pi R_0^3, ~~M^* = \frac{C^*}{2G} R_0^3, 
~~E= - k \left(\frac{R_0}{a_0} \right)^2  \nonumber 
\end{equation}
where $R_0 = r_0 a_0 / (1 + {\textstyle{1\over4}} k r_0^2)$.  Here, the constant $E$ 
is the energy per unit physical mass. So, unlike
the vacuum bulk scenario, a radiative bulk with matter gives rise to two mass terms for
the brane black hole : one, 
$M$ is the physical mass, i.e., the total energy per proper star
volume, and another, $M^*$ is the  `induced mass' of the star that encodes the
bulk information on the brane.  Precisely,
the role of bulk matter is to provide an additional \textit{mass} to the 
collapsing star, resulting in an effective mass 
$M^{\text{eff}} = M + M^*$. It should be noted that this induced mass correction
is quite small due to the presence of $G$. That the physical mass of a braneworld object
is modified  in presence of bulk matter has been shown in \cite{effmass}
and has been depicted as the comoving mass in order to study its cosmological consequences.

We shall now investigate whether there is a static vacuum exterior that 
can match the interior metric  (\ref{int})  
at the boundary of the collapsing star.
The field for the exterior of the collapsing star  is a solution 
of the modified Einstein equation (\ref{eee}) for vacuum  
($T_{\mu \nu} = 0 = {\cal S}_{\mu \nu}$). By employing the RS fine-tuning, the equation reads
\begin{equation}
R_{\mu\nu}=  - {\cal E}_{\mu\nu} + {\cal F}_{\mu\nu} - \frac{1}{2} g_{\mu\nu} {\cal F}
\label{vacu}
\end{equation}
where ${\cal F}$ is the trace of ${\cal F}_{\mu\nu}$. 
We express the  static, spherically symmetric metric for the vacuum exterior by
\begin{equation}
ds^2_{\text{ext}} = -F^2 \left[1- \frac{2Gm}{R}\right]  dT^2 
+\frac{dR^2}{1- \frac{2Gm}{R}} +R^2d\Omega^2
\label{ext}
\end{equation}
where $F = F(R)$ and $m = m(R)$ are, in general, radial functions
characterizing the exterior spacetime given by the coordinates $(T, ~R, ~\Theta, ~\Phi)$.
The matching conditions across the boundary, assumed to satisfy 4D Israel junction conditions, are : 
(i) continuity of the metric and
(ii) continuity of the extrinsic curvature, implying continuity of $\dot R$.
The two conditions are simultaneously satisfied by expressing the metrics 
(\ref{int}) and (\ref{ext}) in terms of null coordinates by adapting the 
method of \cite{coll1}, that results in the following conclusions :
$F(R)=1$ by rescaling, and 
\begin{equation}
m(R) = M + M^* + \frac{3M^2}{8\pi\lambda R^3}  
\label{mass1}
\end{equation}
In terms of the effective mass, the expression for $m(R)$
 can be re-written as
\begin{equation}
m(R) = M^{\text{eff}}  + \frac{3(M^{\text{eff}})^2}{8\pi\lambda R^3}
\label{mass2} 
\end{equation}
In obtaining Eq (\ref{mass2}) from Eq (\ref{mass1}), we have neglected  
higher order terms involving 
$M^*$, since the correction due to the induced mass  is
 sufficiently small so far as the low energy condition $\lambda \gg M/R^3$ holds good. 
The most striking feature of 
the above equation is that it gives  nothing but the Schwarzschild solution for the exterior,
with the  correction term arising from perturbative braneworld gravity.
 However, there is some intriguing difference of the above solution
with the standard 4D Schwarzschild solution. Here the Schwarzschild mass of the star
is not its physical mass $M$  but an effective mass $M^{\text{eff}}$
that incorporates the effects of the radiative bulk.
In the  low energy regime ($\lambda \gg M/R^3$),
this Schwarzschild metric essentially means static exterior. 
Thus, we find that  the exterior may appear to be static even if
there is energy-exchange between the collapsing star and the bulk.
In this sense, the exterior is manifestly static. 

It is worthwhile to make some comments on the gravitational potential of
the collapsing sphere. In braneworlds, the law of gravitation gets modified
from the standard inverse square law, though the correction term becomes 
negligible at low energy \cite{gravpot}.
Eq (\ref{mass2}) reveals that this radiative  bulk
scenario  is in accord with the high energy correction of the Newtonian potential
in braneworld gravity.  The only point is that the correction to the potential 
is further suppressed due to the presence of the bulk matter.

How does the bulk matter play such a crucial role? The answer lies in the
calculation of  the brane Ricci scalar, once by using 
equations (\ref{ext}) and (\ref{mass1}), that gives
\begin{eqnarray}
R^\mu{}_\mu= \frac{9GM^2}{2\pi\lambda R^6}
\label{ricci1}
\end{eqnarray}
and once more by utilizing the trace-free property of ${\cal E}_{\mu \nu}$ 
in the vacuum field equation  (\ref{vacu}). Using the expression for
${\cal F_{\mu\nu}}$ from Eq (\ref{fmunu}), it results in
\begin{equation}
R^\mu{}_\mu =  - \frac{8}{3} ~\kappa_5^2  \psi
\label{ricci2}
\end{equation}
The difference of the above result with the matter-free bulk scenario
is noteworthy. For matter-free bulk, with ${\cal F_{\mu\nu}} = 0$, 
the Ricci scalar in Eq (\ref{ricci2}) had to vanish, that was in direct
contradiction with Eq (\ref{ricci1}), resulting
in the no-go theorem. In the light of the generalized bulk scenario,
this no-go theorem can now be re-stated as : {\em In a braneworld
with vacuum bulk, a collapsing dust sphere cannot have a static exterior.}
On contrary,  for a general non-empty bulk, Eq (\ref{ricci2}) contains a term
comprising of the bulk matter projected onto the brane. A
comparison between Eq (\ref{ricci1}) and  (\ref{ricci2}) reveals that 
a  static  exterior for a collapsing star on the brane
is indeed possible   if the following relation holds good  
\begin{equation}
 \psi = - \left(\frac{3M}{8\pi}  \right)^2 \sqrt{\frac{3 \pi G}{\lambda}} \frac{1}{R^6}
\label{cond}
\end{equation}
where we have used the relations  $\kappa_5^4 = 6 \kappa^2/\lambda$ and $\kappa^2 = 8 \pi G$.
This is exactly what is expected from phantom radiation in the bulk, for which $\psi < 0$.
Thus, the possibility of having a static exterior is quite consistent
with the phantom nature of the bulk field.
In this way,  we arrive at  the radically novel conclusion :
{\em For a general, non-vacuum bulk, a collapsing star on the brane can exchange
energy with the bulk  even if the exterior may appear to be  static to
a braneworld observer.}

Eq (\ref{cond}) in turn gives an idea about the evolution of the 
scale factor. The solution for $\psi$ as obtained from this equation
should satisfy the evolution equation (\ref{eqrho*}) for $\rho^*$, which means 
\begin{equation}
 2 \epsilon \psi - \frac{2 \kappa_5^2}{3 \kappa^2} \left[\dot \psi + 3 \frac{\dot a}{a} \psi \right]
= C^* \frac{\dot a}{a^4}
\label{eqpsi} 
\end{equation}
Together with the expression for $\psi$, it shows the dependence
of  the comoving radius on the brane proper time as
\begin{equation}
C_1~R^3 + C_2~\ln R = C - C_3~\tau
\end{equation}
where the coefficients are given by 
\begin{equation}
C_1 = \frac{M^*}{3} \left(\frac{R_0}{a_0}\right)^3, ~~~
C_2 = \left(\frac{3M}{2\pi}  \right)^2 \frac{3G}{2 \lambda}, ~~~
C_3 =  \epsilon \left(\frac{3M}{2\pi}  \right)^2 \sqrt{\frac{3 \pi G}{\lambda}} \nonumber
\end{equation}
and $C$ is a constant related to $a_0$, the scale factor when the 
star starts collapsing.
The term $C_2$ being too small, it  immediately reveals that
the scale factor for the collapsing star approximately evolves as
\begin{equation}
a^3(\tau) \approx  a_0^3 - B ~\tau
\label{scalefac}
\end{equation}
where the constant $B$ is defined as
\begin{equation}
B =  \frac{6 \epsilon}{M^* }\left(\frac{3M a_0^3}{8\pi R_0^3}  \right)^2 
\sqrt{\frac{3 \pi G}{\lambda}} \nonumber
\end{equation}
Eq (\ref{scalefac}) represents the scale factor of a collapsing star if the constant $B$ is
positive implying $\epsilon >0$. The positive signature for  $\epsilon $ reveals that
the star, in fact, receives phantom radiation from the bulk. 
This provides a possible explanation from bulk viewpoint. The energy loss by the bulk is   
manifest on the brane via the induced mass. 
It is obvious that after a sufficiently long time $\tau \approx a_0^3/B$, 
the scale factor becomes practically zero, so that from the point of view 
of a local observer near the star's surface, the collapse is completed. Further, the scenario 
leads to a naked singularity since the effective mass becomes negative after this time is reached.
 However, this will happen only if the pressure inside the star
remains negligible throughout the process. In this context, it may be noted
that a star with dust in a radiative bulk collapses at a slower rate than a star
with surface pressure in empty bulk \cite{coll3}.


\section{Bulk matter and bulk metric near the brane}

The collapsing  nature of the  star on the brane is manifested by
its motion in the bulk. To a bulk-based observer, the contraction of the star is 
identical to  its motion along the radial direction
of the bulk black hole, with its scale factor  being identified
with the radial trajectory ${\cal R}(\tau)$  at the brane location, with the brane proper time $\tau$
chosen as the parameter \cite{langrev, excos1, graviton, time}.

It should be mentioned here that the global bulk geometry is difficult to obtain
because of the problems associated with embedding simultaneously an FRW brane governing the interior
metric and a static brane region for the exterior of the star, that will result in
different  extrinsic curvatures for the two regions. Consequently, the global bulk metric may
not be strictly VAdS$_{5}$. A possible extension is to  introduce a 
dynamic Swiss-cheese like structure consisting of black cigars penetrating
an FRW brane \cite{scheese}. 
A detailed calculation in this topic is required in order to find out the global bulk metric. 
However, if we focus on the physics near the collapsing region of the brane, then the bulk metric
can be well approximated to be VAdS$_{5}$.
Thus,  one can keep aside the detailed analysis 
and can safely consider the bulk  to be VAdS$_{5}$ locally, so far
as the vicinity of the collapsing region of the brane is concerned.

 This  bulk VAdS$_{5}$ metric is a solution of the 5D field 
equation with the energy-momentum tensor for a radiation field 
\begin{equation}
T_{M N}^{\text{bulk}} =   \psi q_M q_N
\label{bulkem}
\end{equation}
where  $M,N,...$ are the bulk indices and $q_M$ are the outgoing null vectors indicating the energy flow from
the bulk to the brane. Further, for phantom radiation in the bulk, in which we are interested,
the quantity $\psi$ is negative.
With the help of Eq (\ref{cond}), we find that the bulk matter behaves as
\begin{equation}
T_{M N}^{\text{bulk}} = - \left(\frac{3M}{8\pi}  \right)^2 \sqrt{\frac{3 \pi G}{\lambda}}
\left(\frac{a_0}{R_0} \right)^6 \frac{1}{a^6(\tau)}~q_M q_N  
\end{equation}
 Thus, the bulk energy-momentum tensor turns out to be
\begin{equation}
T_{M N}^{\text{bulk}} =  - \left(\frac{3M}{8\pi}  \right)^2 \sqrt{\frac{3 \pi G}{\lambda}}
\left(\frac{a_0}{R_0} \right)^6 \frac{1}{{\cal{R}}^6}~q_M q_N 
\label{bulkmat}  
\end{equation}
The above equation now gives a purely bulk quantity. 
The negative signature  guarantees that  
$T_{M N}^{\text{bulk}}$ represents the energy-momentum tensor of a phantom radiation field.
In this way, we find the nature of the bulk field responsible for the scenario.

One can also find out the bulk metric on the vicinity of the collapsing region of the brane
by the perturbative brane-based approach formulated in \cite{struc}.
In terms of null
coordinate $v = t + \int d {\cal R}/f$, the VAdS$_{5}$ metric, for both incoming
and outgoing radiation, can be written as
\begin{equation}
d S_5^2 = - f({\cal R}, ~v) ~dv^2 - 2  d{\cal R} ~dv + {\cal R}^2 d \Sigma_3^2 
\label{vaid} 
\end{equation}
with $(v, ~{\cal R}, ~x, ~y, ~z)$ representing the bulk coordinates and
  $\Sigma_3$ is the 3-space. The function $f({\cal R}, ~v)$ 
is defined as
\begin{equation}
f({\cal R}, ~v) =  \frac{{\cal R}^2}{l^2} - \frac{{\cal M}(v)}{{\cal R}^2}
\label{fn} 
\end{equation}
with the length scale $l$ related to the bulk (negative) cosmological constant
by $\Lambda_5 = -6 / l^2$ and the mass function
${\cal M}(v)$, the resultant of  the black hole mass $m_1(v)$  and the mass
of the radiation field $q(v)$, is given by
\begin{equation}
{\cal M}(v) = 2 m_1(v) - \frac{q^2(v)}{{\cal R}^2} 
\end{equation}
Now, the on-brane mass function ${\cal M}(\tau)$ is related to the scale factor 
\cite{struc} via
\begin{equation}
{\cal M}(\tau) = \frac{\kappa^2}{3} C(\tau) \propto a(\tau)
\label{bulkmass1}
\end{equation}
By re-definition of parameters involved in  the scale factor of Eq (\ref{scalefac}), 
this function turns out to be
\begin{equation}
{\cal M}(\tau) = {\cal M}_0 (\tau_0 - \tau)^{1/3} 
\label{mtau}
\end{equation}
where ${\cal M}_0 = C_0 B \kappa^2/3$ is its value when the star starts collapsing.
However, this on-brane function will not suffice in describing the bulk geometry
relevant to a braneworld observer.  
We have to further find out the  off-brane mass function ${\cal M}(v)$. 
This can be substantiated by keeping note of the fact that
the function $f({\cal R},~v)$ at the brane-location reduces to 
\begin{equation}
f({\cal R}, ~v)|_{\text{brane}} = \frac{{\cal R}^2}{l^2} 
-  C_0 \frac{{\cal R}(\tau)|_{\text{brane}}}{{\cal R}^2}
= \frac{{\cal R}^2}{l^2} -  C_0 \frac{a(\tau)}{{\cal R}^2}
\end{equation}
The time function for the bulk is, in general, a function of the brane proper time
$t = t(\tau)$. 
With a suitable gauge choice similar to \cite{time}, the bulk time is identical to
the brane proper time, barring an insignificant constant.
By utilizing the identity of the bulk time to the brane proper time, the
null coordinate $v$ turns out to be
\begin{equation}
v = t + \frac{l}{2 \sqrt {\cal M}_0 (t_0 - t)^{1/6}} \left[\frac{1}{2} 
\ln \left(\frac{{\cal R}/l - \sqrt {\cal M}_0 (t_0 - t)^{1/6}}{{\cal R}/l +
\sqrt {\cal M}_0 (t_0 - t)^{1/6}}\right) 
+ \tan^{-1} \frac{{\cal R}/l}{\sqrt {\cal M}_0 (t_0 - t)^{1/6}} \right] 
\label{v} 
\end{equation}

The off-brane mass  ${\cal M}(v)$ at the vicinity of the brane 
can be found out by expanding ${\cal M}(v)$ in
Taylor series around its on-brane value ${\cal M}(t)$ as
\begin{eqnarray}
{\cal M}(v) =  {\cal M}(t) + \left[\frac{\partial {\cal M}}{\partial t_1}\right]_{t_1 = t} 
\int \frac{d {\cal R}}{f}
+\frac{1}{2} \left[\frac{\partial^2 {\cal M}}{\partial t_1^2}\right]_{t_1 = t}
\left(\int \frac{d {\cal R}}{f} \right)^2 + .....
\label{taylor} 
\end{eqnarray}

Hence the off-brane mass function ${\cal M}(v)$ is approximately given by
\begin{equation}
{\cal M}(v) \approx {\cal M}_0 (t_0 -t)^{1/3} + \frac{l}{3} \sqrt {\cal M}_0 (t_0 - t)^{-2/3}
\left[\frac{1}{2} \ln \left(\frac{{\cal R}/l - \sqrt {\cal M}_0 (t_0 - t)^{1/6}}
{{\cal R}/l + \sqrt {\cal M}_0 (t_0 - t)^{1/6}}\right) 
+  \tan^{-1} \frac{{\cal R}/l}{\sqrt {\cal M}_0 (t_0 - t)^{1/6}} \right] 
\label{mv} 
\end{equation}

Thus, by finding out the  null coordinate $v$ and the function ${\cal M}(v)$,
we have been able to obtain the bulk geometry near the collapsing region of the brane, which is sufficient
from the point of view of a braneworld observer.


\section{Summary and open issues}

In this article, we have shown that a generalized non-empty bulk
may lead to a manifestly static exterior for a collapsing spherical star on
the brane. In this scenario, the bulk geometry is  described by a radiative
Vaidya black hole. We have demonstrated that a suitable choice of 
bulk matter can result in the  energy-flow from the bulk to the 
collapsing star  in such a way that a braneworld observer
located at the exterior of the star cannot feel the energy-exchange. 
We have shown, with the help of the brane Ricci scalar,
 that a nontrivial contribution from bulk field is responsible for the scenario.
The scale factor for such a collapsing sphere has also been obtained.
We have also found out the bulk matter 
and derived the bulk geometry near the brane by a brane-based perturbative analysis.
In this way, the article gives a complete picture of both brane and bulk
viewpoints for the situation to be described.

Here we see that the star absorbs energy in course of its collapse. 
It may so happen that 
after a finite time, the star will collect sufficient amount of energy to acquire
pressure. Then its evolution will be governed by the Raychaudhuri
equation (\ref{rc}) with pressure, which might be a generalization of 
the work reported in \cite{coll3} for the non-empty bulk. 
There is another possibility that the star will anti-evaporate.
This situation is somewhat comparable to  
\cite{antievap}. In either case, we need a better understanding of the
bulk-brane dynamics in order to comment further on the ultimate
fate of the collapsing star.

Finally, it is natural to ask : How do we account for  black holes  in this
radiative bulk scenario?  We recall that black holes are not well-understood
even in the empty bulk scenario. Given the situation of the modified conservation
equation for $\rho^*$,  the effective conservation equation for a braneworld black hole
in empty bulk
described in \cite{tidal} will now be modified by a non-trivial contribution
from bulk matter. We expect that the bulk matter contribution may lead to a 
solution for black holes, consistent with the bulk-brane dynamics.
Even the possibility of having  a Schwarzschild solution with an effective mass
discussed in the present article, may be investigated.
However, an extensive study in this field is required for a conclusive remark.

\section*{Acknowledgments}

I am indebted to Sayan Kar for bringing my interest to  braneworld physics
and for a careful reading of the manuscript.
My sincere thanks to Narayan Banerjee, Soumitra Sengupta and Somnath Bharadwaj for their immense help
during the last few years, and to Ratna Koley for  encouragement. 
I also sincerely acknowledge the valuable suggestions of the anonymous referee, which  helped in considerable improvement
of the paper.



\begin{references}

 \bibitem{rs2}  L. Randall and R. Sundrum, Phys. Rev. Lett. {\bf 83} (1999) 4690 
\bibitem{blstring} A. Chamblin, S. W. Hawking and H. S. Reall, 
Phys. Rev. {\bf D61} (2000) 065007
\bibitem{tidal} N. Dadhich, R. Maartens, P. Papadopoulos and V. Rezania,
Phys. Lett. \textbf{B487} (2000) 1
\bibitem{seahra} S. Seahra,  Phys. Rev. {\bf D71} (2005) 084020
\bibitem{adsbh} C. Galfard, C. Germani and A. Ishibashi, Phys. Rev. {\bf D73} (2006) 064014
\bibitem{chargerot} A.  N. Aliev and  A. E. Gumrukcuoglu, Phys. Rev. {\bf D71} (2005) 104027;
\bibitem{revbh} A. S. Majumdar and N. Mukherjee, Int. J. Mod. Phys. {\bf D14} (2005) 1095
\bibitem{bhrecent} S. Creek, R. Gregory, P. Kanti and B. Mistry, hep-th/0606006;
 A. L. Fitzpatrick, L. Randall and T. Wiseman, hep-th/0608208
\bibitem{os} J. R. Oppenheimer and H. Snyder, Phys. Rev. {\bf 56} (1939) 455
\bibitem{coll1} M. Bruni, C. Germani and R. Maartens, Phys. Rev. Lett. \textbf{87}
(2001) 231302
\bibitem{coll2} N. Dadhich and S. G. Ghosh, Phys. Lett. {\bf B518} (2001) 1;
M. Govender and N. Dadhich, Phys. Lett. \textbf{B538} (2002) 233
\bibitem{collgen} G. Kofinas and E. Papantonopoulos, JCAP {\bf 12} (2004) 011
\bibitem{coll3} L. A. Gergely, hep-th/0603254

\bibitem{bulkst} H. S. Reall, Phys. Rev. {\bf D59} (1999) 103506;
J. E. Lidsey, Phys. Rev. {\bf D64} (2001) 063507
\bibitem{bulkph} K. Maeda and D. Wands, Phys. Rev. {\bf D62} (2000) 124009;
A. Mennim and R. A. Battye, Class. Quant. Grav. \textbf{18} (2001) 2171
\bibitem{goldwise} W. D. Goldberger and M. B. Wise,  Phys. Rev. Lett. {\bf 83} (1999) 4922
\bibitem{topo} R. Gregory, Phys. Rev. Lett. \textbf{84} (2000) 2564;
I. Olasagasti and A. Vilenkin, Phys. Rev. {\bf D62} (2000) 044014
\bibitem{coscon} N. Arkani-Hamed, S. Dimopoulos N. Kaloper and R. Sundrum,
Phys. Lett. \textbf{B480} (2000) 193;
C. Csaki, J. Erlich and C. Grojean, Nucl. Phys. \textbf{B604} (2001) 312
\bibitem{adshor} P. Kanti and K. Tamvakis, Phys. Rev. {\bf D65} (2002) 084010
\bibitem{langrev} D. Langlois, Prog. Theor. Phys. Suppl. {\bf 148} (2003) 181
\bibitem{maartbulk} E. Leeper, R. Maartens and C. Sopuerta, Class. Quantum
Grav. {\bf 21} (2004) 1125
\bibitem{lang1} D. Langlois, L. Sorbo and M. Rodriguez-Martinez, Phys. Rev. Lett. {\bf 89} (2002) 171301;
D. Langlois and L. Sorbo, Phys. Rev. {\bf D68} (2003) 084006; 
D. Langlois, Astrophys. Space Sci. 283 (2003) 469
\bibitem{chamb}  A. Chamblin, A. Karch and A. Nayeri, Phys. Lett. {\bf B509} (2001) 163

\bibitem{vaid} P. C. Vaidya, Nature {\bf 171} (1953) 260;
C. W. Misner, Phys. Rev. {\bf 137} (1965) B1360;
R. W. Lindquist, R. A. Schwartz and C. W. Misner, Phys. Rev. {\bf 137} (1965) B1364
\bibitem{maartrev} R. Maartens, Living Rev. Relativity {\bf 7} (2004) 7 

\bibitem{vads5gen} L. A. Gergely, Phys. Rev. {\bf D68} (2003) 124011
\bibitem{asymgen}  L. A. Gergely, E. Leeper and R. Maartens, Phys. Rev. {\bf D70} (2004) 104025;
I. R. Vernon and D. Jennings, JCAP {\bf 07} (2005) 011
\bibitem{nonradial} D. Langlois, Prog. Theor. Phys. Suppl. {\bf 163} (2006) 258
\bibitem{radbh5} D. Jennings, I. R. Vernon, A. C. Davis and C. van de Bruck, 
JCAP {\bf 04} (2005) 013;
L. A. Gergely and Z. Keresztes, JCAP {\bf 01} (2006) 022;
A. R. Frey and A. Maharana,  JHEP {\bf 08} (2006) 021

\bibitem{phant1} R. R. Cadwell, M. Kamionkowski and N. N. Weinberg, Phys. Rev. Lett. {\bf 91} (2003) 071301;
 S. M. Carroll, M. Hoffman and M. Trodden, Phys. Rev. {\bf D68} (2003) 023509;
 P.H. Frampton,  Phys. Lett. {\bf B555} (2003) 139;
L. Perivolaropoulos, J. Cosmol. Astropart. Phys. {\bf 10} (2005) 001;
S. Tsujikawa, Phys.Rev. {\bf D72} (2005) 083512;
E. J. Copeland, M. Sami and S. Tsujikawa, hep-th/0603057
\bibitem{phant2} S. Capozziello, S. Nojiri and S. D. Odintsov, Phys. Lett. {\bf B632} (2006) 597
\bibitem{phantworm} S. A. Hayward,  Phys. Rev. {\bf D65} (2002) 124016;
L. A. Gergely,  Phys. Rev. {\bf D65} (2002) 127503;
H. Koyama and S. A. Hayward,  Phys. Rev. {\bf D70} (2004) 084001
\bibitem{rkskphant} R. Koley and S. Kar, Mod. Phys. Lett. {\bf A20} (2005) 363
\bibitem{rksk6d} R. Koley and S. Kar, Class. Quant. Grav. {\bf 24} (2007) 79

\bibitem{eee} T. Shiromizu, K. Maeda and M. Sasaki, Phys. Rev. {\bf D62} (2000) 024012;
A. N. Aliev and A. E. Gumrukcuoglu, Class. Quant. Grav. {\bf 21} (2004) 5081 
\bibitem{struc} S. Pal, Phys. Rev. {\bf D74} (2006) 024005

\bibitem{maeda} K. Maeda, Lect. Notes Phys. {\bf 646} (2004) 323
\bibitem{excos1} D. Langlois and M. Rodriguez-Martinez, Phys. Rev. {\bf D64} (2001)
123507
\bibitem{excos2} G. Kofinas, G. Panotopoulos and T. N. Tomaras, JHEP {\bf 01} (2006) 107
\bibitem{excos3} I. Brevik, J. Mattis Borven and S. Ng, Gen. Rel. Grav. {\bf 38} (2006) 907
\bibitem{dynscal} S. Mizuno, K. Maeda and K. Yamamoto, Phys. Rev. {\bf D67} (2003) 023516
\bibitem{effmass} P. S. Apostolopoulos and N. Tetradis, Phys. Rev. {\bf D71} (2005) 043506;
P. S. Apostolopoulos, N. Brouzakis, E. N. Saridakis and N. Tetradis, Phys. Rev. {\bf D72} (2005) 044013;
P. S. Apostolopoulos and N. Tetradis, Phys. Lett. {\bf B633} (2006) 409
\bibitem{gravpot} M. Sasaki, T. Shiromizu and K. Maeda, Phys. Rev. {\bf D62} (2000) 024008; 
J. Garriga and T. Tanaka, Phys. Rev. Lett. \textbf{84} (2000) 2778

\bibitem{graviton} C. Cartier, R. Durrer and M. Ruser, Phys. Rev. {\bf D72} (2005) 104018
\bibitem{time} J. Garriga and M. Sasaki,  Phys. Rev. \textbf{D62} (2000) 043523
\bibitem{scheese} L. A. Gergely, Phys. Rev. {\bf D74} (2006) 024002
\bibitem{antievap} R. Casadio and C. Germani, Prog. Theor. Phys. {\bf 114} (2005) 23;
R. Casadio and C. Germani, J. Phys. Conf. Ser. {\bf 33} (2006) 434


\end{references}
\end{document}